# Enhanced stability of 2D organic-inorganic halide perovskites by doping and heterostructure engineering

Rahul Singh,[1] Prashant Singh[2] and Ganesh Balasubramanian[3]

[1]Department of Mechanical Engineering, Iowa State University, Ames, IA 50011

[2]Division of Materials Science & Engineering, Ames Laboratory, Ames, Iowa 50011, USA

[3]Department of Mechanical Engineering & Mechanics, Lehigh University, Bethlehem, PA 18015

**Abstract**

Organic-inorganic halide perovskite solar cells have recently attracted much attention due to their low-cost fabrication, flexibility, and high-power conversion efficiency. The reduction from three- to two-dimension (2D) promises an exciting opportunity to tune the electronic properties of organic-inorganic halide perovskites. Here, we propose first-principles density-functional theory based route to study the effect of reduced dimensionality, impurity doping, and heterostructure engineering on energy stability, band-gap and transport properties of 2D hybrid organic-inorganic halide perovskites. We show that the energetic stability of two-dimensional organic-inorganic halide perovskites can be significantly enhanced by chemically depositing $MoS_2$ monolayer as a precursor in the system by heterostructure engineering. While on one hand, the structures have similar and excellent transport properties as their bulk counterparts, on the other, they possess the advantage of a broad range of tunable band gaps. Our predictions will expedite future efforts (both theoretical and experimental) in the synthesis, measurements and applications of high performance perovskites.

Over the last decade, perovskites have been one of the most intensively examined classes of materials because of their outstanding optoelectronic properties.[1-6] The versatility of these materials encompasses a series of optoelectronic devices such as light-emitting diodes, [1,2] transistors,[3] lasing applications[4], as well as other intriguing electronic properties.[5–11] In particular, organic-inorganic halide perovskites (OIHP), archetypically $CH_3NH_3PbX_3$ (X = Cl, Br or I), have attracted significant attention because of their remarkable photovoltaic properties,[12–14] achieving power conversion efficiency as high as 22.1%.[4] Thin OIHP films that are typically synthesized and examined for the electronic transport, suffer from stability issues. A potential solution is to employ planar heterostructures in lieu of the 3-dimensional (3D i.e., bulk) films.[13,15–23] Quantum confinement effects in 2D chemistries increase the band gap due to a blue shift that decreases with increasing number of layers in a heterostructured material.[24–27] While increasing the number of inorganic layers decreases the bandgap, simultaneously it is detrimental to the stability.

Renewed interest in thermoelectrics is motivated by the realization that complexity at multiple length scales can lead to new mechanisms for high *performance perovskite* materials. Theoretical predictions suggested that the thermoelectric efficiency could be greatly enhanced by quantum confinement[28] by ionic doping or heterostructure engineering.[29-30] The heterovalent or isovalent metal ions such as ($Sn^{2+}$, $Cd^{2+}$, $Zn^{2+}$, $Mn2+$)[31][32–34] have been introduced as dopants with the possibility of imparting paramagnetism to increase stability. The heterostructure engineering by chemical layer deposition has also been tried to enhance the photoluminescence,[35] charge-transfer mechanism,[35] and surface dopant.[36] Both doping and chemical layer deposition can lead to quantum confinement (QC) of charges. The QC helps narrowing down electron energy bands with the decreasing dimensionality, which produces high effective masses and Seebeck coefficients. Moreover, similar sized heterostructures decouple the Seebeck coefficient and electrical conductivity by electron filtering[37] that could result in improved performance. Despite all the progress, our theoretical understanding of doping and heterostructure engineering effect on underlying electronic and transport properties of perovskites is sparse and limited.

We employ first-principles density functional theory (DFT) to examine the stability, electronic-structure and transport properties of [001] terminated 2D OIHPs.[38] To test our hypothesis that ionic doping and inclusion of a $MoS_2$ML precursor can improve stability of OIHPs, we stack 2D assemblies of $MAPbI_3$ on top of $MoS_2$ML by constraining them within the interaction distances. Our results reveal that a single sheet of $MoS_2$ML considerably improves the power factor (*PF*) of OIHPs. The enhanced stability of the hybrid OIHPs is explained via the structural and electronic properties. In addition, we investigate the effect of magnetic impurity ($Mn^{2+}$) doping on thermoelectric (TE) properties of 2D $MAPbI_3$ heterostructures, as a potential route to modulate the optical properties of halide perovskites.

We examine bulk and 2D variants of $MA(Pb/Sn)I_3$ OIHPs using first-principles density functional theory (DFT)[39,40] based Vienna ab initio simulation package (VASP).[41–43] We construct [001] terminated 2D OIHPs from 3D $MA(Pb/Sn)I_3$. For geometry optimization and electronic structure calculations, we use the projected augmented-wave (PAW) basis[44] and the Perdew–Burke–Ernzerhof (PBE)[41] exchange-correlation functional. The charge and forces are converged to $10^{-5}$ eV and 0.01eV/Å, respectively, using energy cut-off of 800 eV. The Monkhorst-Pack[44] k-mesh grid of 7×7×3, and 3×3×5 is used for (2D $MAPbI_3$, 2D $MASnI_3$, Mn-doped-$MAPbI_3$) and $MAPbI_3$ML/$MoS_2$ML, respectively. The tetrahedron method with Blöchl corrections is used to calculate the density of states (DOS).[41] The thermoelectric properties are calculated using the BoltzTrap[45] code interfaced with VASP.

*Structural analysis of bulk and 2D variants of $MAXI_3$ (X=Pb, Sn):* Hybrid organic-inorganic perovskites have an $ABX_3$ architecture, where A is a monovalent organic cation, $CH_3NH_3^+$ (i.e., $MA^+$), while B is a metal cation (i.e., $Pb^{2+}$, $Sn^{2+}$), and X is a halide anion (i.e., Cl , Br , I or their mixtures). In typical perovskite crystals, B occupies the center of an octahedral $[BX_6]^{46}$ cluster, while A is 12-fold cuboctahedrally coordinated with X anions.[15,47] Similar to previous studies, we consider the high temperature pseudo-cubic phase of $MAPbI_3$.[48,49] The calculated (experimental)[50,51] lattice constants of bulk $MAPbI_3$ are $a = 6.432$ (6.361) Å, $b = 6.516$ (6.361) Å, $c=6.446$ (6.361) Å, and α≈β≈γ≈90°. The predictions indicate a 3.41% increase in the simulated equilibrium cell volume relative to that in the experiments,[50,51] with 1.27% increase in average bond length of Pb-I atoms. For 2D MAPbI$_3$, the calculated (experimental) lattice parameters of the PbI$_2$ surface are: a = 6.437 (6.361) Å and b = 6.449 (6.361) Å, respectively. We find a 1.38% increase in lattice constant in [110] plane of the simulated materials, although the thickness of the slab shrinks by 2.5%. The reduced thickness (Fig. 1 (c)) of the slab shrinks the overall slab volume by 1.33% with respect to experimental bulk MAPbI$_3$.

In case of Mn-doped 2D MAPbI$_3$, the calculated (experimental[48,49]) lattice parameters are: a = 6.361 (6.361) Å and b = 6.445 (6.361) Å, respectively. The doped structure shows 25% distortion in Mn-I polyhedra, while only 9% distortion in Pb-I polyhedra. The average Mn-I bond-length increases by 4% relative to Pb-I. However, for MoS$_2$ML/MAPbI$_3$ML, a = 6.362 (6.361[48,49]) Å and b = 6.428 (6.361[48,49]) Å, respectively, similar to that of bulk MAPbI$_3$. The average Pb-I bond length is 3.201 Å similar to the value obtained in the bulk MAPbI$_3$. Also, MoS$_2$ML/MAPbI$_3$ML surface has negligible PbI$_2$ polyhedral distortion, which suggests of the structural stability of MAPbI$_3$ML "on surface disposition". The structural differences caused by MoS$_2$ML enhances the stability of the system as noted from the relative energies of the six structures shown in Fig. 2.

*Electronic properties of bulk and 2D variants of MAXI$_3$ (X=Pb, Sn):* Experimentally, the bandgap of cubic MAPbI$_3$ is ~ 1.5-1.62 eV.[7] Our standard PBE calculations predict that bulk MAPbI$_3$ is a semiconductor with a direct bandgap of 1.726 eV, which agrees reasonably with the experiments. Using PBE electronic structure predictions as a reference, we perform additional band structure calculation of 2D MAPbI$_3$. The results indicate that both surfaces have a direct bandgap with the conduction band minimum (CBM) and valence band maximum (VBM) located at the M point of the Brillouin zone, as illustrated in Fig. 3. The energy states ranging from -3 to 3 eV are mostly contributed by Pb and I atoms, signifying their influence in the physical and chemical properties of these perovskites. Specifically, the states near the top of the valence band are predominated by I-*5p* and Pb-*6s* states, while the conduction bands are constituted by Pb-*6p* states with hybridization of I-*5p* states. As shown earlier in Fig.1, the surface terminated by PbI$_2$ contains more Pb atoms, which leads to a broader conduction band and thus a narrower bandgap of the 2D MAPbI$_3$ terminated by PbI$_2$.

The spin-polarized band structure for 10% Mn-doped 2D MAPbI$_3$ are presented in Fig. 4 (a). The bandgap of the Mn-doped OIHP at Pb-site is 1.2 eV, while Mn at Sn site reduces the bandgap to ~1 eV (below, we discuss only Mn-doping at Pb-site). Both the top of the valence band and the bottom of the conduction band are located at the M [½ ½ 0] point. The doping results in a more stable structure by distorting the neighboring environment, see charge density plot in Fig. 4 (b) (top), compared to pure 2D structures (Fig. 2). However, the bandgap of Mn-doped OIHP decreases due to large structural distortion. The Mn spin-up channel is completely filled, and down-spin channel is completely empty leading to strong magnetic behavior (Mn-moment ~ 5 $\mu_B$) as shown by magnetization density plot in Fig. 4 (b) (bottom). The top flat band just below the Fermi level, corresponds to the Mn impurity band, which is attributed to the Mn-*3d* state. The Mn-doping changes the electronic structure near the Fermi level affecting stability as well as the transport properties. While there is no notable improvement in transport properties, the stability of the structure is enhanced by Mn-doping.

Next, we investigate the effect of charge-transfer interactions in MoS$_2$ML/MAPbI$_3$ML. The adsorption of organic molecules is expected to modify the electronic properties of MoS$_2$ML. The MoS$_2$ML isolated layer has a predicted direct bandgap of 1.73 eV, closer to the experimentally observed observation band-gap of 1.70 eV.[52,53] The VBM of MoS$_2$ML is contributed by Mo-*4d* and S-*3p* states, while the CBM is mainly contributed by Mo-*4d* states and feebly by S-*3p* states.

To obtain a deeper insight into the electronic properties of the now functionalized MoS$_2$ML, we compute the density of states (DOS) and full charge densities (projected and full, which includes VBM and CBM) for MoS$_2$ML/MAPbI$_3$ML, illustrated in Figure 5. The adsorption of MAPbI$_3$ML introduces new flat energy levels in the bandgap region of MoS$_2$ML resulting in a bandgap of ~1.27 eV. As the interactions between MAPbI$_3$ML and MoS$_2$ML are weak, the band structure of functionalized MoS$_2$ML is effectively a combination of those of MoS$_2$ML and the adsorbed MAPbI$_3$ML. Hence, the bandgap reduction is attributed to the recombination of the energy levels. Nevertheless, for MAPbI$_3$ML functionalized MoS$_2$ML, the new energy levels appear in the region of the conduction band, indicating that MoS$_2$ML can be tuned into a p-type semiconductor by doping with MAPbI$_3$ML. As shown in Fig. 5 (right-panel), for MoS$_2$ML/MAPbI$_3$ML both CBM and VBM are solely contributed by MAPbI$_3$ML. In the band-structure plot, MoS$_2$ predominantly lies below -0.2 eV with respect to the Fermi level, E$_F$. The conduction bands with an energy range from ~1 - 3 eV are derived largely from the Pb-*s/p* and I-*s* states. The valence bands with a range from -3 eV – 0 eV exert strong hybridization between the Pb-*p*, I-*p*, Mo-*d* and S-*p* states. The localized distribution of electron density on MAPbI$_3$ML and MoS$_2$ML indicates negligible effect of the interlayer interactions but is responsible for higher stability of MoS$_2$ML/MAPbI$_3$ML as compared to the bulk structure. The weaker interaction, in Fig. 5 (c), is shown by small band-distortion at the interface of MoS$_2$ML and MAPbI$_3$ML, where bond-angle between Pb-I-Pb changes from 180º to ~179º. We also note from this structural analysis that band-distortions are reduced to ~5% in MoS$_2$ML/MAPbI$_3$ML compared to ~10-20% in MAPbI$_3$ML or Mn-doped MAPbI$_3$ML.

*Thermoelectric properties of bulk and 2D variants of MAXI$_3$ (X=Pb, Sn):* A band-structure possessing a large-effective mass in CBM, with a minimum of about $5k_BT$ above the VBM, can potentially achieve nonmonotonic Seebeck coefficient, and hence a large *PF*.[54] Deposition of MAPbI$_3$ML on MoS$_2$ML introduces resonant states that can achieve a tailored band structure. We determine the band effective mass, *m\** from second order derivative of the band energy with respect to the wave vectors:

$$\left(\frac{1}{m^*}\right) = \frac{1}{\hbar^2} \frac{\partial^2 E_n}{\partial k_x \partial k_y}$$

where, *x* and *y* are the directions in reciprocal space, *n* is the band index, $E_n$ is the band energy, and $\hbar$ is the modified Planck's constant. The derivatives have been evaluated at CBM for electrons (*m$^e$*) and at VBM for holes (*m$^h$*). The reduction in dimensions from bulk to 2D creates opportunities to optimize the highly anisotropic structures due to the smaller directional effective mass (electrical conductivity, $\sigma \propto \frac{1}{m^*}$ ).[55]

The calculated effective hole/electron mass for bulk MAPbI$_3$ is $(0.17; 0.19; 0.28)m_h^*$ and $(0.27; 1.51; 0.12)\ m_e^*$, respectively. The three values in the parentheses represent the three perpendicular directions (x, y, z), while for 2D structures we consider the two planar directions (x, y). For 2D MAPbI$_3$, the corresponding values are $(0.17; 0.33)m_h^*$ and $(1.53; 0.85)\ m_e^*$, respectively. The calculated effective mass for 2D MAPbI$_3$ indicates an obvious in-plane anisotropy in effective mass due to [100] stacking. The strong directional effective mass is larger along [100] stacking, and the direction orthogonal to the stacking [010] predicts smaller values. Thus, we deduce that the carrier mobility is high (small effective mass) along [010], with heavy mass states present in the transverse direction, i.e., [100]. Similarly, the Mn-doped 2D MAPbI$_3$ shows large anisotropy in effective mass along [100] stacking at Pb-site [$(0.12; 5.42)\ m_h^*$, $(2.50; 1.15)\ m_e^*$]. In contrast to the Mn-doped 2D MAPbI$_3$, the energy minima/maxima shift to the X [½ 0 0] point from the M point in MoS$_2$ML/MAPbI$_3$ML. The calculated hole- and electron-effective masses at M-point are $m_h^* = (0.76; 1.13)$ and $m_e^* = (3.40; 1.51)$, respectively. The large anisotropy in the effective mass along the [100] direction facilitates hole transport holes along [100] compared to the electrons. However, along the [010] direction, both electrons and holes experience equal effective mass resulting in similar transport properties.

In TE materials, a high Seebeck coefficient (*S*) at a given carrier concentration results from a high overall DOS effective mass ($m_d^*$). However, $\sigma$ decreases with increasing $m_d^*$, and also depends on the inertial effective mass, *m*\*. Ioffe showed empirically that for doped semiconductors to be good TEs, the sweet-spot for carrier concentration, $n \sim 10^{18}$–$10^{20}$ per cm$^3$, corresponding to degenerate semiconductors or semimetals.[56] Here, we predict a similar range for $n_{hole}$ and $n_{electron}$. As the doping concentration increases, $\sigma$ increases and *S* decreases. Figs. 6 (a) and (b) present *S* predictions for 2D MAPbI$_3$ (left-panel) and MoS$_2$ML/MAPbI$_3$ML (right-panel). In comparison with bulk MAPbI$_3$,[57] we find that *S* of 2D MAPbI$_3$ is ~20% lower.[57] The large *S* in bulk MAPbI$_3$ arises from the DOS and band mobility. As shown earlier in Fig. 3, the DOS above and below the Fermi level have a significant contrast in the bulk structures[53] compared to the 2D counterparts, and the smaller *m*\* in 3D materials leads to an increased band mobility[57] that is important to enhance the electronic transport. In Fig. 6, we show that bulk MAPbI$_3$ has slightly better

TE properties than 2D MAPbI$_3$ and MoS$_2$ML/MAPbI$_3$ML, however, 2D counterparts have better energetic stability (see Fig. 2). For all practical purposes, the latter is more important, as stability of materials signifies that 2D MAPbI$_3$/MoS$_2$ML/MAPbI$_3$ML can be formed in a laboratory environment and will not degrade as quickly as bulk MAPbI$_3$.[47]

$PF = S^2\sigma$ of a TE material quantifies their electrical power generation ability. The most efficient TE materials possess high $\sigma$ and high $S$. Since $\sigma$ and $S$ have competing dependencies on $n$ simultaneously obtaining high values for both properties is challenging. We compare the calculated $PF$ per relaxation time as a function of temperature for MAPbI$_3$ML and that with MoS$_2$ML. Although both materials originate from the parent cubic crystal structure, differences are noted in their TE properties (Figs. 6(a)–(f)). The dissimilarities in $\sigma$ of 2D MAPbI$_3$ and MoS$_2$ML/MAPbI$_3$ML originates from latter's intrinsic $sp$-type bonding. The 2D OIHP deposited on MoS$_2$ has slightly lower $PF \sim 29.3 \times 10^{10}$ W/m.K$^2$.s relative to 2D MAPbI$_3$ ($\sim 32.5 \times 10^{10}$ W/m.K$^2$.s) due to an enhanced thermal electronic conductivity.[57] Both 2D MAPbI$_3$ and MoS$_2$ML/MAPbI$_3$ML show strong temperature dependence in $S$, but $\sigma$ essentially remains temperature invariant. Thus, the $PF$ predictions for specific chemical potential value at 800 K, Fig. 6 (e) and (f), are ~ 100% and ~ 200% higher than at 500 and 300 K, respectively, for both materials. Note that $PF$ of 2D MAPbI$_3$ is ~10% higher than MoS$_2$ML/MAPbI$_3$ML, but an enhanced energetic stability of the latter contributes to its selection for potential TE applications. On one hand, OIHPs possess the advantage of hybridization, while on the other hand, they demonstrate the ability to tune TE properties through nanostructuring. In addition, the $PF$ can be further tuned by controlling the ionic doping concentrations.

In this paper, we propose approaches to manipulate the electronic and thermoelectric properties of 2D OIHPs. From first-principles calculations, we show that energetic stability of 2D OIHPs and its variants can be tuned by impurity doping and by monolayer deposition. Our findings show that the chemically deposited monolayer of 2D MAPbI$_3$ on MoS$_2$ provides optimal bandgap required for photovoltaic applications. The enhanced bandgap, compared to 2D MAPbI$_3$, arises from the weak interactions between MoS$_2$ and MAPbI$_3$ monolayers. Moreover, the weak interlayer interactions energetically stabilize MoS$_2$ML/MAPbI$_3$ML. The MoS$_2$ML/MAPbI$_3$ML also possesses the excellent thermoelectric properties as noted in bulk MAPbI$_3$. We believe our predictions provide a useful guideline for future experiments examining the stability of OIHP based solar materials and suggest a practicable approach to enhance the light-harvesting capability of MoS$_2$ML/MAPbI$_3$ML.

**Acknowledgment**

The research was supported, in part, by the National Science Foundation (NSF) grant no. CMMI-1404938. The work at Ames Laboratory was supported by the U.S. Department of Energy (DOE), Office of Science, Basic Energy Sciences, Materials Science & Engineering Division, which is operated by Iowa State University for the U.S. DOE under contract DE-AC02-07CH11358.

**References**

(1) Jaramillo-Quintero, O. A.; Sanchez, R. S.; Rincon, M.; Mora-Sero, I. Bright Visible-Infrared Light Emitting Diodes Based on Hybrid Halide Perovskite with Spiro-OMeTAD as a Hole-Injecting Layer. *J. Phys. Chem. Lett.* **2015**, *6* (10), 1883–1890.

(2) Tan, Z.-K.; Moghaddam, R. S.; Lai, M. L.; Docampo, P.; Higler, R.; Deschler, F.; Price, M.; Sadhanala, A.; Pazos, L. M.; Credgington, D.; et al. Bright Light-Emitting Diodes Based on Organometal Halide Perovskite. *Nat. Nanotechnol.* **2014**, *9* (9), 687–692.

(3) Chin, X. Y.; Cortecchia, D.; Yin, J.; Bruno, A.; Soci, C. Lead Iodide Perovskite Light-Emitting Field-Effect Transistor. *Nat. Commun.* **2015**, *6* (1).

(4) Xing, G.; Mathews, N.; Sun, S.; Lim, S. S.; Lam, Y. M.; Gratzel, M.; Mhaisalkar, S.; Sum, T. C. Long-Range Balanced Electron- and Hole-Transport Lengths in Organic-Inorganic CH3NH3PbI3. *Science* **2013**, *342* (6156), 344–347.

(5) Jung, H. S.; Park, N.-G. Perovskite Solar Cells: From Materials to Devices. *Small* **2015**, *11* (1), 10–25.

(6) Kojima, A.; Teshima, K.; Shirai, Y.; Miyasaka, T. Organometal Halide Perovskites as Visible-Light Sensitizers for Photovoltaic Cells. *J. Am. Chem. Soc.* **2009**, *131* (17), 6050–6051.

(7) Ogomi, Y.; Morita, A.; Tsukamoto, S.; Saitho, T.; Fujikawa, N.; Shen, Q.; Toyoda, T.; Yoshino, K.; Pandey, S. S.; Ma, T.; et al. $CH_3NH_3Sn_xPb_{(1-x)}I_3$ Perovskite Solar Cells Covering up to 1060 Nm. *J. Phys. Chem. Lett.* **2014**, *5* (6), 1004–1011.

(8) Snaith, H. J. Perovskites: The Emergence of a New Era for Low-Cost, High-Efficiency Solar Cells. *J. Phys. Chem. Lett.* **2013**, *4* (21), 3623–3630.

(9) Song, T.-B.; Chen, Q.; Zhou, H.; Jiang, C.; Wang, H.-H.; (Michael) Yang, Y.; Liu, Y.; You, J.; Yang, Y. Perovskite Solar Cells: Film Formation and Properties. *J. Mater. Chem. A* **2015**, *3* (17), 9032–9050.

(10) Sum, T. C.; Mathews, N. Advancements in Perovskite Solar Cells: Photophysics behind the Photovoltaics. *Energy Env. Sci* **2014**, *7* (8), 2518–2534.

(11) Weber, D. CH3NH3PbX3, ein Pb(II)-System mit kubischer Perowskitstruktur / CH3NH3PbX3, a Pb(II)-System with Cubic Perovskite Structure. *Z. Für Naturforschung B* **1978**, *33* (12).
10

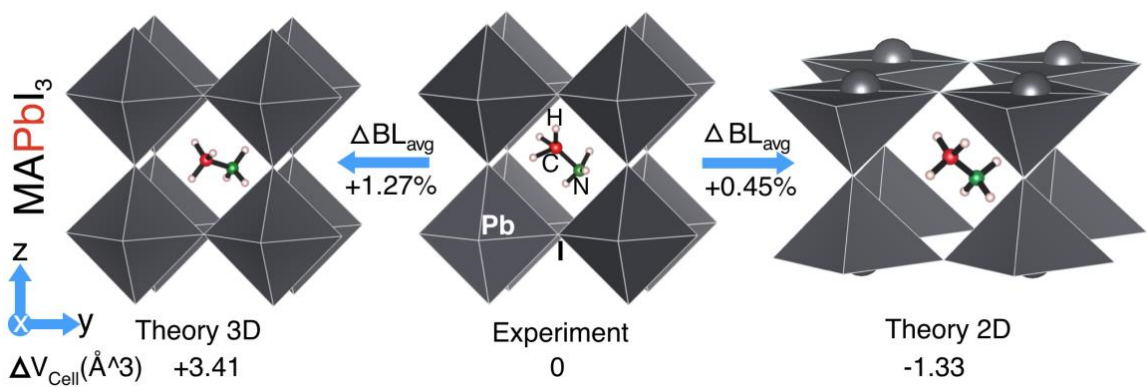

**Figure 1:** (Color online) The side view of bulk MAPbI$_3$ unit cell as obtained from (a) simulations and (b) experiments. The side view of (c) (001) terminated 2D MAPbI$_3$ unit cell. The unit cell volume of the bulk and energetically relaxed structure increases by 3.41% relative to the experimentally measured volume, while the 2D cell shrinks by 1.33%. In comparison to the experiments the average Pb-I bond-length of bulk and 2D MAPbI$_3$ increases by 1.27 and 0.45%, respectively. For the relaxed 2D MASnI$_3$, the unit cell volume and average Sn-I bond length increases by 1.65 and 2.05%, respectively, compared to bulk (see supplement).

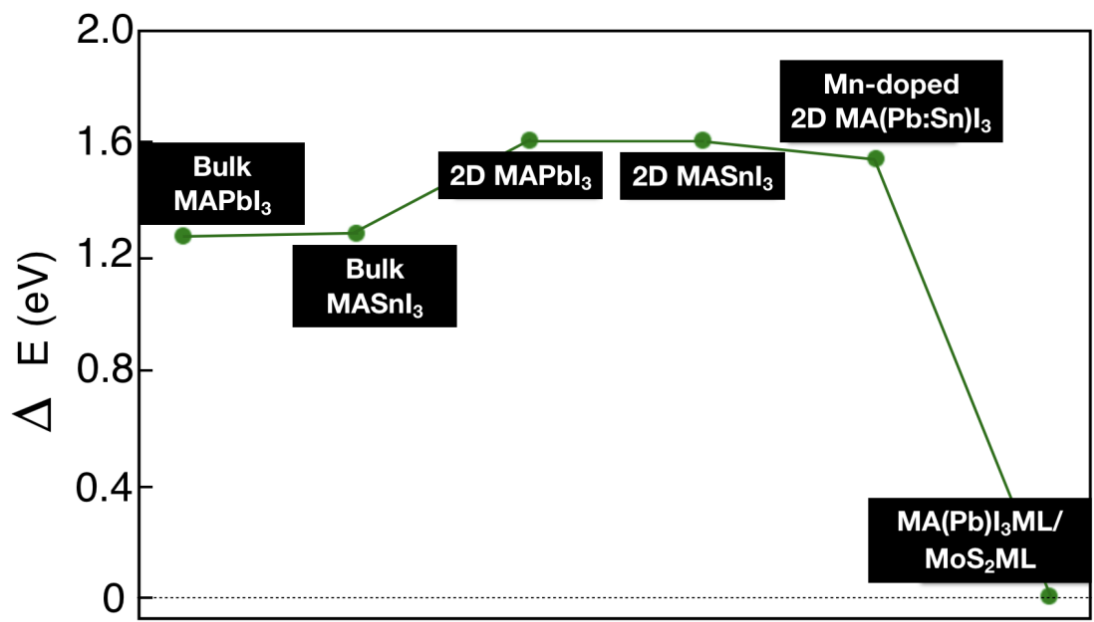

**Figure 2:** (Color online) Relative formation energies (eV) of bulk, 2D, Mn-doped and MAPbI$_3$ML/MoS$_2$ML MAPbI$_3$ suggest that the MAPbI$_3$ML/MoS$_2$ML offers the highest structural stability (lowest formation energy). ML stands for monolayer.

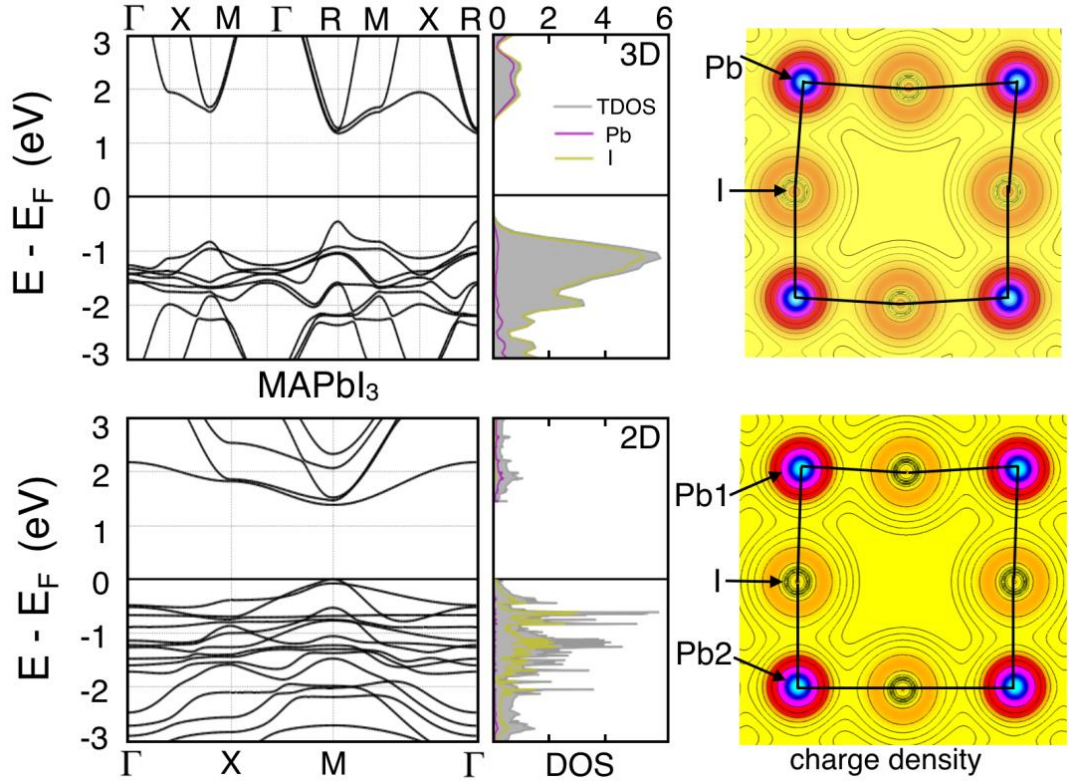

**Figure 3:** (Color online) Comparative band-structure, density of states (DOS) and (001) projected charge density of bulk and [001] terminated 2D $MAPbI_3$. We can see that going from bulk to 2D makes DOS more structured and bands flatter. Both bulk and 2D $MAPbI_3$ show calculated direct band-gap of 1.726 eV and 1.27 eV at R-point and M-point of Brillouin zone, respectively. The [001] projected charge density of 2D $MAPbI_3$ reveals enhanced bonding between Pb and I (as seen by non-circular lobes both at Pb and I-sites) compared to bulk. Identical isosurface values are employed for the charge density plots.

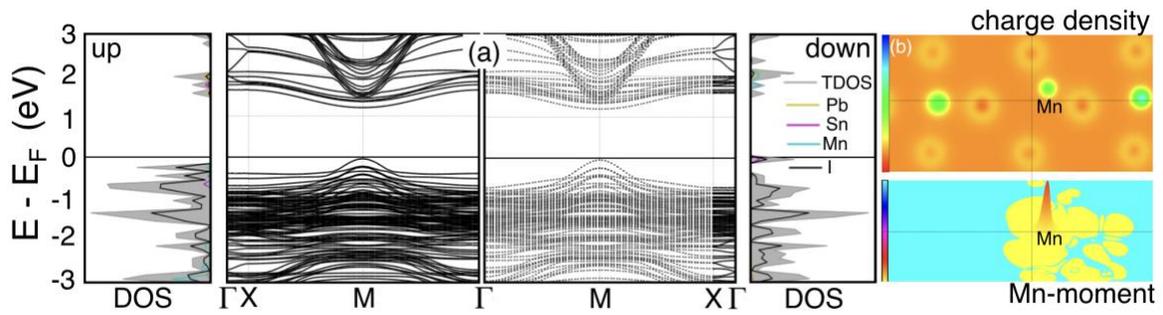

**Figure 4:** (Color online) (a) Spin-polarized band structure, (b) charge density (top) and magnetization density (bottom) of Mn-doped (at Pb-site) 2D MAPbI$_3$ monolayer stacked along [010] direction. In Mn-doped MA(Pb;Sn)I$_3$, the strong magnetic characteristic of Mn with high saturation magnetization of ~5 Bohr magneton ($\mu_B$), leads to induced $I$ moments on nearby sites by distortion of the MnI$_2$ polyhedra. Mn-doping stabilizes MAPbI$_3$ML almost by ~0.12 eV with respect to 2D MAPbI$_3$ and 2D MASnI$_3$ (see Fig. 2).

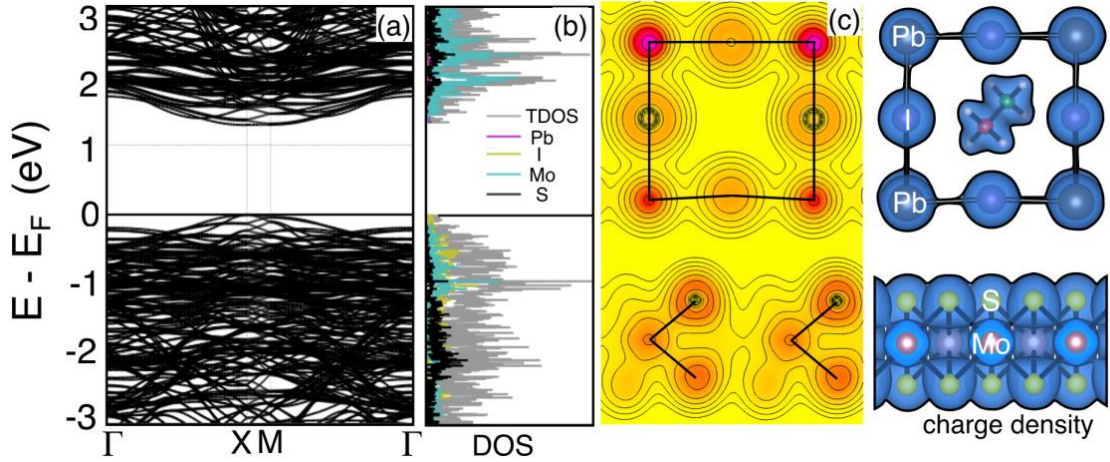

**Figure 5:** (Color online) (a) The electronic band-structure, (b) density of states (center), and (c) (110) projected and full charge density of chemically deposited MAPbI$_3$ monolayer (ML) on MoS$_2$ML. The MAPbI$_3$ML/MoS$_2$ML enhances the band gap to 1.27 eV with respect to Mn-doped MAPbI$_3$ML (see Figs. 4). Chemical layer deposition of MAPbI$_3$ML on MoS$_2$ML enhances energetic stability compared to bulk, 2D, and Mn-doped MA(Pb;Sn)I$_3$ML. We attribute increased stability to the weaker interlayer interaction originating from Pb-*6p* and S-*3p* states.

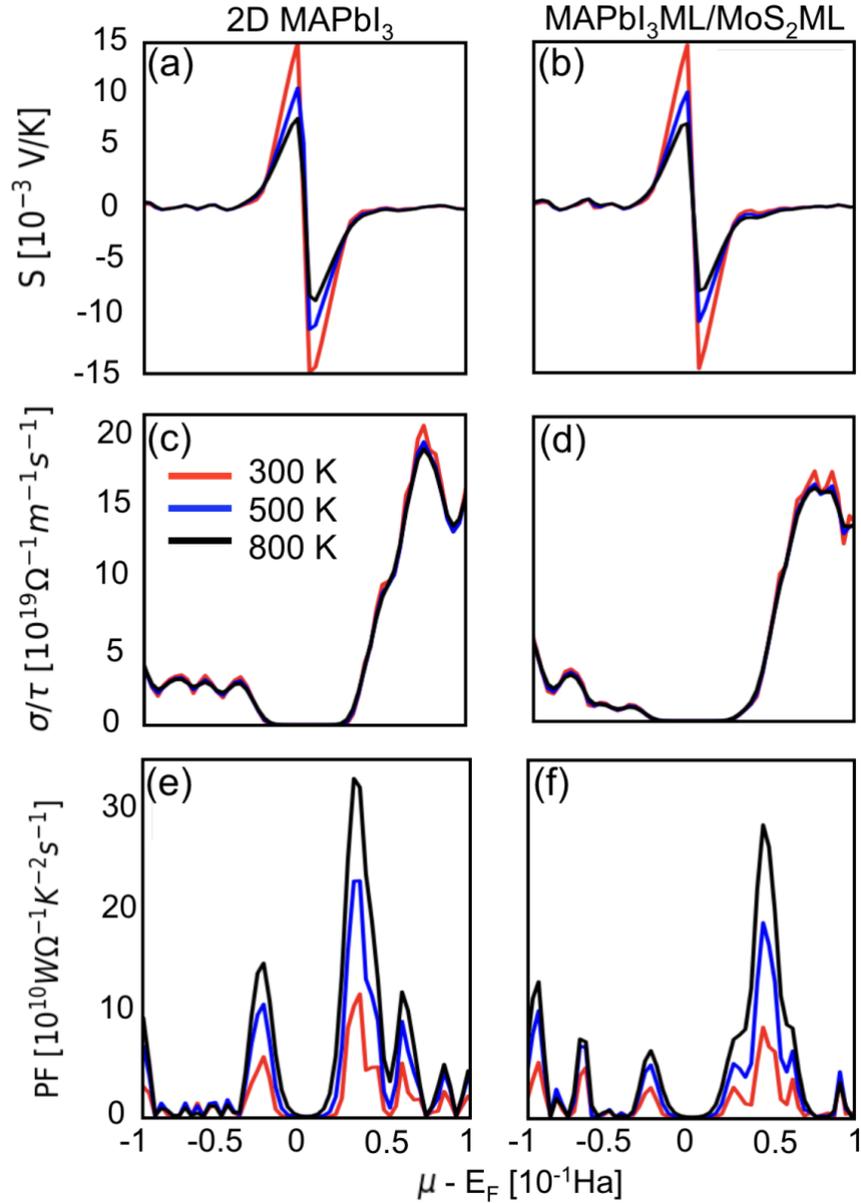

**Figure 6:** (Color online) The Seebeck coefficient (a, b), the electrical conductivity (c, d), and the power factor (e, f) of 2D MAPbI$_3$ (left-panel) and MAPbI$_3$ML/MoS$_2$ML (right panel) are presented at 300 K, 500 K and 800 K, respectively (ML = monolayer). The optimal value of power factor (e, f) for the two cases occur at a chemical potential of +0.025 and +0.04 Hartree, respectively. We show that MAPbI$_3$ML/MoS$_2$ML retain the thermoelectric properties of bulk with an order of magnitude increase in the energy stability (see Fig.2).[57]